\documentclass{owrart}

\usepackage{color}
\usepackage{graphicx}
\usepackage{amsthm}

\begin{document}


\begin{talk}[]{Daniela Cadamuro}
{The massive modular Hamiltonian for a double cone}
{Cadamuro, Daniela}

\noindent

Since it  has been set up in the 1970's due to works by Tomita that became public with lectures by Takesaki~\cite{1970Takesaki}, as well as by Araki~\cite{1976Araki}, Tomita-Takesaki modular theory has been one of the most important developments in the theory of operator algebra, as well as in quantum theory.  However, in relevant examples from quantum (field) theory,  obtaining an ``explicit'' form of the modular generator $\log \Delta$ has been the strenous work of many researchers along the time.  At least in the following situations, a model-independent answer is known:
\begin{itemize}
\item If $\mathcal{M}$ is the algebra of all observables and $\Omega$ represents a thermal equilibrium state (KMS condition), then $\log\Delta$ is the generator of time translations (up to a factor) \cite{1967HaagHugenholtzWinnink}.

\item If $\mathcal{M}= \mathcal{A}(\mathcal{W})$ is the algebra associated with a spacelike wedge region $\mathcal{W}$ in quantum field theory, and $\Omega$ is the Minkowski vacuum, then $\log\Delta$ is the generator of boosts along the wedge \cite{1975BisognanoWichmann}.
\end{itemize}
But what about the algebra of a double cone, $\mathcal{M}=\mathcal{A}(\mathcal{O})$, in a quantum field theory?
To answer this question we consider the example of a real scalar free field $\phi$ of mass $m>0$. We consider the Fock vacuum as our cyclic and separating vector. The local algebras, as well as the modular operator \cite{EckmannOsterwalder1973}, are determined by second quantization,  so that we only need to consider the modular operator at one-particle level, which is defined as follows.

On the (complex) one-particle Hilbert space $\mathcal{H}_1$ of the theory, we consider a (closed real) local subspace $\mathcal{L}_1(\mathcal{O}) \subset \mathcal{H}_1$, which is ``standard'' and ``factorial'' ($\overline{\mathcal{L}_1 +i\mathcal{L}_1}=\mathcal{H}_1$, $\mathcal{L}_1 \cap \mathcal{L}'_1= \mathcal{L}_1 \cap i\mathcal{L}_1 = \{ 0\}$), where ``prime'' denotes the symplectic complement. 
We define the one-particle Tomita operator on $\mathcal{H}_1$ as 
\begin{equation}
T_1 \, :\, f +ig \mapsto f-ig, \quad f,g \in \mathcal{L}_1(\mathcal{O}),
\end{equation}
the polar decomposition of its closure is $T_1 = J_1 \Delta_1^{1/2}$. (We shall drop the index ``1'' from now on.)

We can rewrite the one-particle modular generator as follows. Let $P$ be the real-linear projector onto $\mathcal{L}\subset \mathcal{H}$ with kernel $\mathcal{L}' \subset \mathcal{H}$. Then, on a certain domain, we write
\begin{equation}
P= (1+T)(1-\Delta)^{-1}.
\end{equation}
A computation then shows that
\begin{align}\label{FGformula}
 \log\Delta = -2 \operatorname{arcoth} (P -iPi -1).
\end{align}
This determines $\log\Delta$ from $P$, and hence from $\mathcal{L}$ \cite{1989FiglioliniGuido}. We now write this formula in a different manner, by writing $\mathcal{H}$ in time-0 formalism in configuration space. Here, $\mathcal{H}$ is parametrized by time-0 initial data of field and field momentum $f = (f_+,f_-)$. The scalar product and the complex structure, with $A= -\nabla^2 + m^2$, are given by
\begin{equation}
\operatorname{Re}\langle f,g \rangle_{\mathcal{H}}= \Big\langle f, 
 \begin{pmatrix}
A^{1/2} & 0 \\
0 & A^{-1/2} 
\end{pmatrix} g\Big\rangle_2, \quad i_A =
 \begin{pmatrix}
0 & A^{-1/2} \\
-A^{1/2} & 0
\end{pmatrix}.
\end{equation}
The local subspaces are defined as follows. Let $\mathcal{B}$ be the base of $\mathcal{O}$ at time 0, then we define
\begin{equation}
\mathcal{L} = \overline{\mathcal{C}^\infty_0(\mathcal{B}) \oplus \mathcal{C}^\infty_0(\mathcal{B})}, \quad P=\chi \oplus \chi,
\end{equation}
where $\chi$ multiplies with the characteristic function of $\mathcal{B}$. Inserting this in formula \eqref{FGformula}, we have
\begin{equation}
\log \Delta = i_A 
\begin{pmatrix}
0 & M_- \\
-M_+ & 0
\end{pmatrix},
\end{equation}
where
\begin{equation}\label{eq:m}
M_\pm = 2A^{\pm \frac{1}{4}}\operatorname{arcoth}(B)A^{\pm \frac{1}{4}}, \quad B= \overline{A^{1/4}\chi A^{-1/4}} + \overline{A^{-1/4}\chi A^{1/4}} -1.
\end{equation}
Hence, $\Delta$ is determined from $\chi$ and $A$. However, ``explicitly'' finding  the spectral decomposition of $B$ as a selfadjoint operator on $L^2(\mathbb{R}^s)$ is very difficult. There are however known examples:
\begin{itemize}
\item If $\mathcal{O}$ is the wedge in $x_1$-direction, then $M_-$ multiplies with $2\pi x_1$, indepedent of $m$.

\item If $\mathcal{O}$ is a double cone of radius $r$ and $m=0$, then $M_-$ multiplies with $\pi(r^2 - \|\pmb{x} \|^2)$ \cite{1982HislopLongo}.
\end{itemize}
Now the questions we would like to answer in the case of double cones and $m>0$ are the following: Is $M_-$ mass indepedent? Is $M_-$ a multiplication operator? Since answering these questions anaytically is very difficult, we do it numerically, namely we evaluate $B$ and $M_-$ numerically to check this hypothesis.

\begin{figure}
\includegraphics[width=\textwidth]{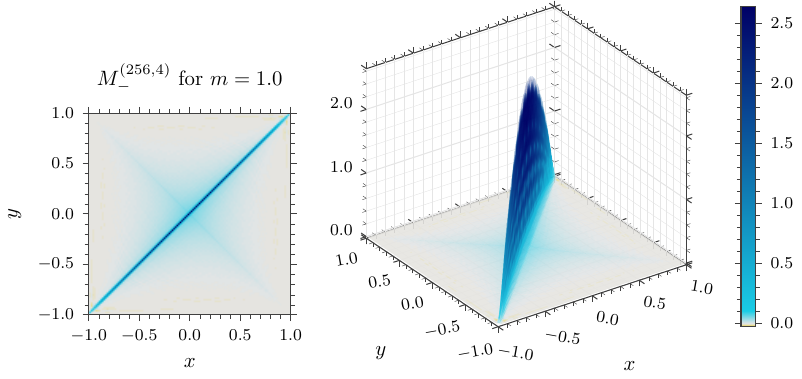}
\caption{}\label{fig:kernelMm}
\end{figure}

Using numerical approximation means approximating $A^s$ and $\chi$ with finite-dimensional matrices. For that, we need to choose an orthonormal basis and finite dimensional in one summand of $\mathcal{H}$, and we need to approximate $A^{\pm 1/4}$ and $\chi$ with a matrix in this basis. Then, we can apply numerical eigendecomposition in order to evaluate the $\operatorname{arcoth}$, and therefore approximate the operator $B$. We do this with no rigorous estimates on the approximation. Explicitly, $A,\chi$ acts on $L^2_{\mathbb{R}}(\mathbb{R})$ by $A= - \partial_x^2 +m^2$, and $\chi$ is determined by the region considered: $\chi(x)= \Theta(x)$ for a wedge, or $\chi(x) = \Theta(1+x)\Theta(1-x)$ for the standard double cone.

As our basis functions, we choose suitable piecewise linear functions \cite{BosCadMin2023}, and the discretization is first done for $A^{-1/4}$ which is bounded and has a known convolution kernel; we then obtain $A^{1/4}$ by numerical matrix inversion. 
We can then approximate (the integral kernel of) $M_-$ using the formula \eqref{eq:m}; this is done by functional calculus of matrices, and the computation turns out to require extended floating point precision of 400--600 decimal digits.  We expect convergence against the undiscretized result in the weak sense, i.e., if $M^{(N,b)}$ denotes the integral kernel at a number $N$ of basis elements covering the interval $[-b,b]$,
 \begin{equation*}
    \iint g(x)  M^{(N,b)}_-(x,y) h(y) dx\, dy \xrightarrow[N,b\to\infty]{} \iint g(x) M_-(x,y) h(y) dx\,dy.  
 \end{equation*}
 We choose $g=h$ to be a Gaussian located near a point $\mu$, then we vary this point $\mu$. 

\begin{figure}
\includegraphics[width=\textwidth]{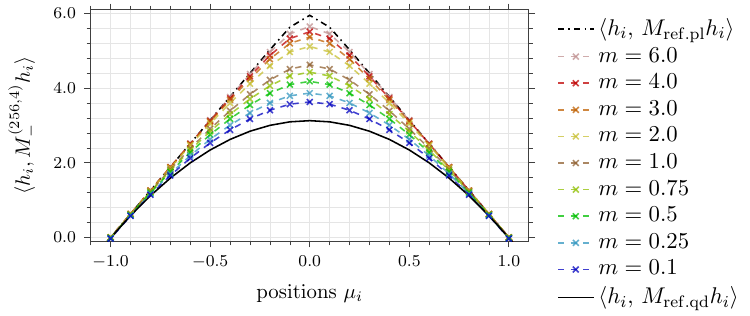}
\caption{}\label{fig:Mm_mass}
\end{figure}

The results in the wedge case turn out to be compatible with known results.  
In the case of a double cone, we find that the discretized kernel $M_-$ is concentrated predominantly on the diagonal, see Figure~\ref{fig:kernelMm}. There appear to be some contributions along the antidiagonal, but it is unclear whether this is due to numerical errors or whether there is really a subdominant non-diagonal contribution. An explicit expression for the curves displayed in Figure~\ref{fig:kernelMm} and Figure~\ref{fig:Mm_mass} is not known. The smeared version of the discretized kernel $M_-$, see Figure~\ref{fig:Mm_mass}, shows that the kernel is mass-dependent. In particular, the black parabola corresponds to the case $m=0$ and therefore to the quadratic result of Hislop-Longo, while the two straight black lines (piecewise linear) for large mass correspond to the result of a left and a right wedge. Indeed, large masses correspond to small correlation lengths, and hence a heuristic explanation for the approximate ``double wedge'' structure may be that at one end of the interval, the contribution from the other end of the interval is very small, so that the modular operator for the interval approximately behaves like the one for a half-line.

A similar analysis can be done for a double cone in the 3+1-dimensional field using its spherical symmetry. It turns out in this case that the modular operator also depends on angular momentum \cite{BosCadMin2023}.

\end{talk}

\end{document}